\documentclass[pof,twocolumn]{revtex4-1}
\usepackage{graphicx}
\usepackage{amssymb,amsmath}

\begin{document}

\title{From linear stability analysis to three-dimensional organisation \\ in an incompressible open cavity flow}

\date{\today}

\author{L.R. Pastur}
\affiliation{Universit\'e Paris Sud 11, F-91405 Orsay Cedex, France}
\affiliation{LIMSI-CNRS BP 133, F-91403 Orsay Cedex, France}
\author{Y. Fraigneau}
\affiliation{LIMSI-CNRS BP 133, F-91403 Orsay Cedex, France}
\author{F. Lusseyran}
\affiliation{LIMSI-CNRS BP 133, F-91403 Orsay Cedex, France}
\author{J. Basley}
\affiliation{Universit\'e Paris Sud 11, F-91405 Orsay Cedex, France}
\affiliation{LIMSI-CNRS BP 133, F-91403 Orsay Cedex, France}
\affiliation{Laboratory For Turbulence Research in Aerospace and Combustion, Department of Mechanical and Aerospace Engineering, Monash University, Clayton 3800, Australia}

\begin{abstract}
 Three-dimensional direct numerical simulations of an incompressible open square cavity flow are conducted. Features of the permanent (non-linear) regime together with the linear stability analysis of a two-dimensional steady base flow are discussed. Spanwise boundary conditions are periodic and control parameters set such that the shear layer is stable against Kelvin-Helmholtz modes. Three branches of destabilising modes are found. The most destabilising branch is associated with steady modes, over a finite range of spanwise wavenumbers. The two other branches provide unsteady modes. Features of each branches are recovered in the permanent regime: wavelength of the most powerful spanwise Fourier mode, swaying phenomenon, angular frequencies, indicating that modes of each branches are selected and interact in the permanent flow. 
 
\end{abstract}
 
\maketitle

\section{Introduction}

Open cavity flows belongs to the family of impinging flows, which are known to develop self-sustained oscillations. When driven by a lid, cavity flows do not exhibit such self-sustained oscillations. Production of tones requires the existence of a shear layer, that forms between the outer driving flow and the inner flow, velocities in the inner flow being one order of magnitude smaller than in the outer flow, in the permanent regime. For decades, an extensive literature has been devoted to the study of self-sustained oscillations occurring in such flows in general, and in open cavity flows in particular. Among the most striking results, one may cite earlier works by Rossiter, who derived a phenomenological law for explaining the frequency selection for the the so-called Rossiter modes, in compressible flows \cite{Ros64}, Gharib and Roshko who made distinction between shear-layer (resonant) modes and wake modes \cite{GhaRos87}, Miksad who demonstrated the amplification and non-linear coupling of modes excited in the free shear-layer \cite{Mik74,MikJonPowKimKha82}, Rockwell and co-authors, who intensively studied experimentally impinging and cavity flows, and identifying several regimes of oscillations \cite{Roc77,RocKni79,RocNau79,KniRoc82}. Among other features, they have explained the occurrence of amplitude modulations, at impingement, in which lower frequencies are involved. Recently, Delprat derived an empirical formulation for explaining spectral features of cavity flow oscillations for various control parameters, in compressible regimes, that may be extended to the incompressible regime \cite{Del06,Del10}. Recent investigations have considered cases of non-linear competition of modes of oscillation, resulting in a so-called mode switching phenomenon, both experimentally \cite{KegSpiGarCat04,PasLusFauFraPetDeb08} and numerically \cite{FarPaaSza12}. In recent studies, coherent structures associated with self-sustained modes have been investigated in experiments by means of modern high-speed particle image velocimetry techniques \cite{Basley2010}.
Our present understanding of open cavity flows, in the incompressible limit, may be roughly summarized as follows. The shear layer that forms between the outer and inner flows is known to be unstable with respect to streamwise Kelvin-Helmholtz modes, beyond some critical value of the Reynolds number. Kelvin-Helmholtz modes develop, downstream in the shear layer, from perturbations initiated at the leading corner. The unstable shear layer rolls up on itself to produce vortices that cyclically collide with the trailing edge. Changes in the pressure field at impingement are instantaneously fed back to the leading edge, due to incompressibility or, equivalently, when the cavity length is small in regard to the acoustic wavelength and the material velocity small before sound velocity. Feedback then initiates the growth of a new perturbation at the leading edge. Therefore, both leading and trailing edges are locked in phase, giving rise to strong self-sustained oscillations, whose frequency both depends on the incoming flow velocity, the cavity length, and the boundary layer thickness at the leading edge. 

What has been far less studied is the inner flow, which exhibits non-trivial three-dimensional organisation  \cite{NeaSte87,BreCol08,Faure2007,Faure2009}. Centrifugal instabilities were shown to give rise, beyond a critical Reynolds number, to Taylor-G\"ortler-like rolls~\cite{Faure2007,Faure2009}. The flow curvature along the trailing wall bends the pathline of material particles, which are subject to inertial forces. When viscous terms cannot overcome centrifugal effects, instabilities may develop, eventually giving rise to the formation of stable raw of vortical rolls, in the spanwise direction, as observed in experiments or direct numerical simulations of the flow \cite{BreCol08,AliRobGlo12,Faure2009}. In experiments, as well as in three-dimensional direct numerical simulations of the flow, where spanwise walls are rigid, rolls drift toward one or the other wall, depending on their initial spanwise location  \cite{Faure2007,Faure2009}. Drift was interpreted as initiated by a pumping effect at the walls. More recently, Alizard \textit{et al} \cite{AliRobGlo12} conducted direct numerical simulations of the cavity flow, with periodic spanwise boundary conditions, in a configuration where the shear layer has become unstable against Kelvin-Helmholtz modes. In this study, a linear stability analysis has been conducted, on the (unstable) steady base flow, with respect to spanwise Fourier modes, and several branches of (linearly) growing modes were found. 

The present contribution relies on the intrinsic features of the inner flow, in an open cavity flow at a Reynolds number below the threshold of Kelvin-Helmholtz instability. Three-dimensional direct numerical simulations of the cavity flow are conducted with periodic spanwise boundary conditions, in order to get rid of the Eckman pumping effect. In the case under study, though the shear layer remains stable against Kelvin-Helmholtz modes, the inner flow exhibits a three-dimensional organisation. Beside the formation of a spanwise raw of pairs of vortical structures, a detailed analysis of the permanent regime shows that the regime is not steady. Unsteadiness comes from a slightly swaying motion of the main structure, very similar to the one observed in \cite{AliRobGlo12}. A linear stability analysis is conducted, on a two-dimensional steady state with respect to three-dimensional perturbations. As it will be shown, three branches of (spanwise) growing modes are found, whose features are recovered in the permanent regime with little distortion, though coupled through non-linear terms.

\section{Cavity flow features}

Instabilities in the shear layer primarily depend on the dimensionless cavity length $\Gamma _{\theta _0}=L/\theta_0$, ratio of cavity length, $L$, to boundary layer momentum thickness, $\theta_0$, at the leading edge. Reynolds numbers based on $L$ and incoming velocity $U_0$ (at the upstream cavity wall), is another relevant control parameter. For a lower part, cross-stream aspect ratio $\Gamma_L = L/H$, defined as the ratio of cavity length, $L$, to cavity height, $H$, is appropriate when considering cavity inner-flow interactions with the flapping shear layer. Finally, Strouhal numbers based on incoming velocity, $U_0$, and $L$, constitute an adequate normalisation for frequencies corresponding to shear layer self-sustained oscillations.

\begin{figure}
  \centering{
  \begin{tabular}{c}
  \includegraphics[width=.75\linewidth]{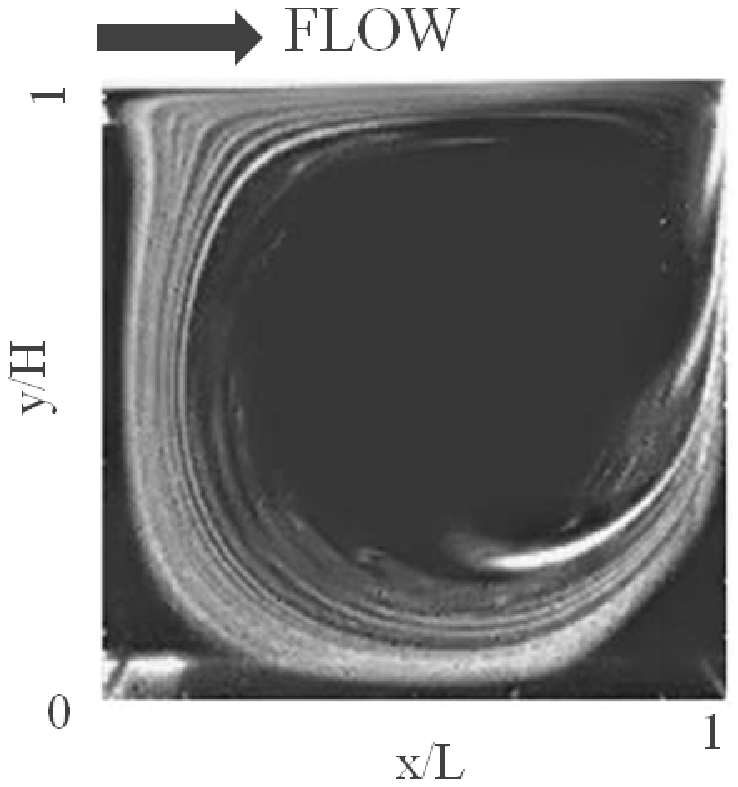} \\
  \includegraphics[width=.75\linewidth]{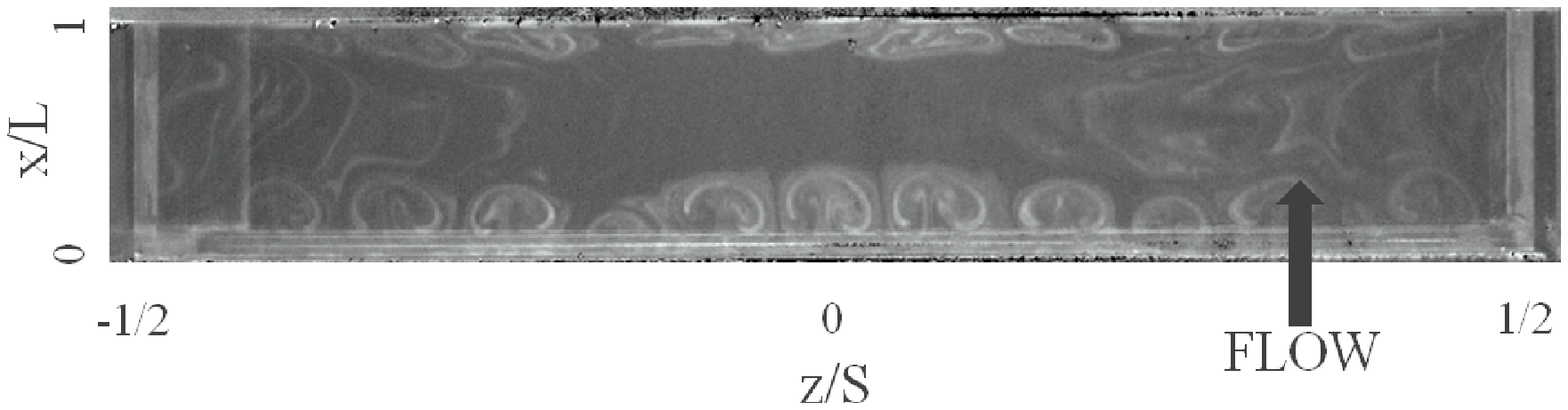} \\
	\end{tabular}
	}	
	\caption{Instantaneous visualisation of the flow seeded with smoke, for $\Gamma _L= 1$,
$\Gamma_S = 6$, $\mathrm{Re}_H=2\,300$. (Top) Cross-stream view in a plane ($x, y$),
(bottom) top-view in a plane ($x, z$), located inside the cavity at $y/H = -0.3$. One can distinguish the shear-layer and the main recirculation in the inner-flow (top) and a (horizontal) spanwise cut of Taylor-G\"ortler rolls revealed by the seeding smoke (bottom).}
	\label{fig:tgvisu}
\end{figure}
\begin{figure}
	\centering{
  \includegraphics[width=.75\linewidth]{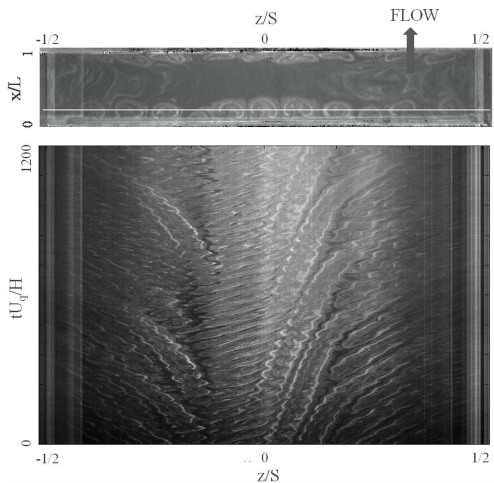}
	}
	\caption{Top: spanwise topview of the flow. Bottom: space-time diagram of the Taylor-G\"ortler rolls dynamics.  The space-time diagram is made of the vertical concatenation of the spatial line shown in the top picture (the line intercepts the mushroom-like structure, close to the cavity upstream wall). From an experimental cavity flow, at $\Gamma _L=1$, $Re_H\simeq 5\,800$.}
	\label{fig:xtplot}
\end{figure}

The curvature induced by the main flow recirculation, inside the cavity, shown in Figure~\ref{fig:tgvisu}a from an experimental flow, is responsible for the development of centrifugal instabilities, over some range of the control parameters. Relevant parameters, there, are the Reynolds number based on cavity height, $Re_H=U_0H/\nu $, and $\Gamma _L$. In \cite{Faure2009}, it was shown that an unambiguous parametrization of centrifugal instabilities requires a third dimensionless number; spanwise ratio $S/H$ was chosen, although $S$, the cavity span, mainly contribute to the selection of number of spanwise structures. Rather, $H/\theta _0$ would be a better choice, although not always easy to estimate in experiments. In the non-linear saturated regime, centrifugal instabilities give rise to vortical structures that form a spanwise alley of pairs of counter-rotating vortices. A spanwise cut of this alley of vortices, in a top view of an experimental flow, is shown in Figure~\ref{fig:tgvisu}b, in a cavity where $\Gamma_L=1$. Each pair is shaped like a torus around the main recirculation flow~\cite{Faure2007,Faure2009}. 
Spanwise walls, at $z/S=\pm 1/2$, generate a B\"odewadt pumping, responsible for a drift of the Taylor-G\"ortler rolls toward one wall or the other, as can be seen in the experimental space-time diagram of Figure~\ref{fig:xtplot}. In this figure, oblique lines correspond to drifting structures, and it is clear, from the different slopes that can be seen, that inner spanwise dynamics is associated with several drift velocities, depending on the spanwise position, $z$.
Henceforth, drift motions introduce additional unsteadiness in the flow, enriching the low-frequency range of power spectra. 

\section{Direct numerical simulations}
\label{sec:dns}

A three-dimensional direct numerical simulation is performed with the cavity aspect ratio set to $\Gamma _L=1$ and Reynolds number to $Re= 3\,850$. Spanwise boundary conditions are periodic, to get the intrinsic features of the inner flow.

\subsection{Numerical methods}

The study addresses an incompressible and isothermal flow whose governing equation can be described by the non-dimensional Navier-Stokes equations:
 \begin{equation}
 \left\{\begin{array}{rcl} 
 \displaystyle \frac{\partial \boldsymbol{U}}{\partial t} + \left(\boldsymbol{U}\cdot \nabla\right)\boldsymbol{U} & = & \displaystyle -\nabla P + \frac{1}{Re}\Delta \boldsymbol{U} \\
 & & \\
 \nabla\cdot \boldsymbol{U} & = & 0 \\
 \end{array}\right.
 \label{eqn:ns}
\end{equation}
where $\boldsymbol{U}$ is the velocity field and $P$ the pressure field. 
Numerical simulations are performed with the OLORIN code developed at LIMSI, which is based on an incremental prediction -- projection method, see \cite{gadoin2001} for more details. Momentum equations are discretised with a finite volume approach on a staggered structured grid. The spatial discretisation of fluxes is carried out with a second-order centred scheme in a conservative form and time derivation is approximated by a second-order backward differentiation formula. Viscous terms are implicitly evaluated whereas convective fluxes are explicitly estimated at time $t^{n+1}$ by means of a linear  Adams-Bashford extrapolation. The discretised form of the Navier-Stokes equations yields a Helmholtz-type problem of the form:
$$ 
 \left\{\begin{array}{rcl} 
 \displaystyle \left(I- \frac{2\Delta t}{3Re}\nabla^{2}\right)\boldsymbol{U}^{n+1} & = & -\nabla P^{n+1}+ S^{n,n-1} \\
 & & \\
 \nabla \cdot \boldsymbol{U}^{n+1} & = & 0 \\
 \end{array}\right.
$$
where superscript $n$ tags time $t_{n}$, $\Delta t$ is the time step and $S^{n,n-1}$ is the source term gathering all explicit quantities, evaluated at times $t_n$ and $t_{n-1}$.
For each time-step, the numerical procedure is splitted in two parts, a prediction step and a projection step. The former consists in resolving the Helmholtz equation by considering the explicit pressure field $P^{n}$ in place of the implicit one. The integration is performed with an ADI (Alternating Direction Implicit) method \cite{kn:hirsch87}. As a result, we obtain an estimated velocity field $\boldsymbol{U}^{*}$ that is not yet divergence-free. The incompressibility property is imposed by using an incremental projection method \cite{kn:Goda1979}. The projection step requires to resolve a Poisson-type equation, using a relaxed Gauss-Seidel method coupled to a multigrid method, in order to accelerate convergence, where the source term relies on non-zero divergence of the predicted velocity field:
$$ 
 \nabla^{2}\phi = \nabla \cdot \boldsymbol{U}^{*}
$$
Solution $\phi$ corresponds to the pressure time-increment, gradient $\nabla \phi $ is the correction term such that the velocity field is divergence free at time $t^{n+1}$. The Poisson equation is commonly solved with Neumann-type boundary conditions, where the normal derivative on the domain limits is zero. By doing so, the boundary condition, on the corresponding normal velocity component, is not affected by the correction term.

\subsection{Geometric setup}

The geometric setup consists of an open cavity capped with a parallelepipedic duct in which is generated the channel flow driving the inner cavity flow. A cartesian coordinate system ($x,y,z$), for streamwise, crosswise and spanwise directions, respectively, is set midspan at the top of the upstream cavity wall. The cavity dimensions are $L=5$~cm, $L/H=1$, $S/H=6$. Upstream and downstream lengths of the duct are respectively $L_u/L=1$ and $L_d/L=3$ and its height is $H_v/H=3$. The total domain is meshed on $160\times128\times192$~nodes, among which $64\times64\times192$ are devoted to the cavity. The mesh, regular in the spanwise direction, is particularly refined close to the walls and at the cavity-top in order to enhance the spatial resolution of boundary layers and shear-layer.

The inlet flow is determined by Dirichlet boundary conditions. In order to limit the numerical domain size, and therefore CPU time-consumption, the upstream vein length has been reduced. The inlet velocity profile is preliminary calculated by means of a 2D simulation of a laminar channel flow in spatial development, representative of the experimental upstream vein. The profile is then extracted out of the appropriate cross-section of the channel-flow and extruded in the periodic spanwise direction. The Reynolds number is set at $Re= 3\,850$ (bulk velocity $U_q=1.2$~m/s), and the boundary layer momentum thickness is $\theta _0/H=26\times 10^{-3}$. Usual non-sliding conditions are applied at the walls. The numerical simulation is carried out over a time-duration of $60$~s (1440 time-units $H/U_q$), after overtaking the numerical transitory state of the flow.

\subsection{Intrinsic features of Taylor-G\"ortler vortices}

\begin{figure}
 \begin{tabular}{c}
	\includegraphics[width=.8\linewidth]{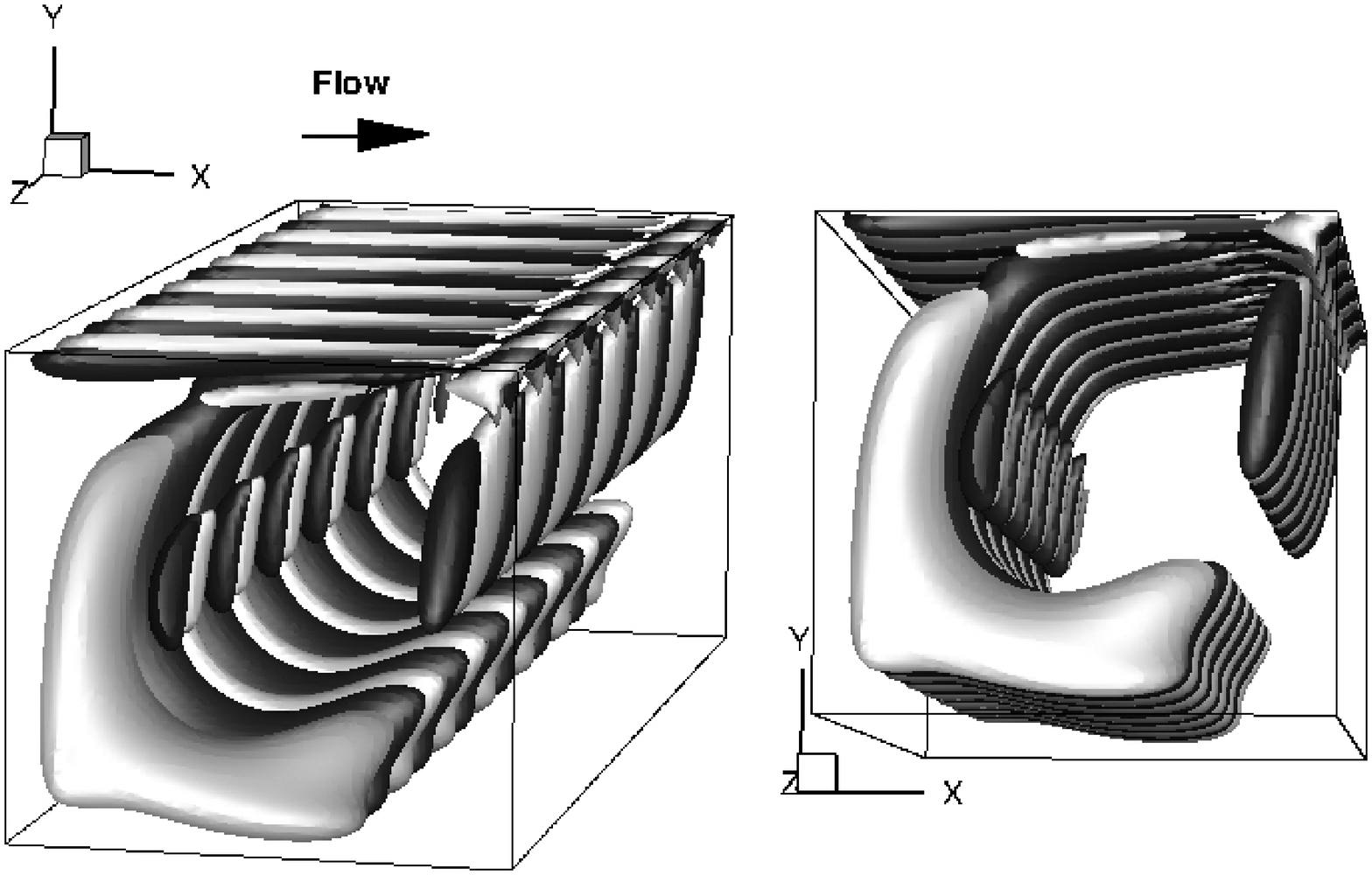} \\
	\includegraphics[width=.8\linewidth]{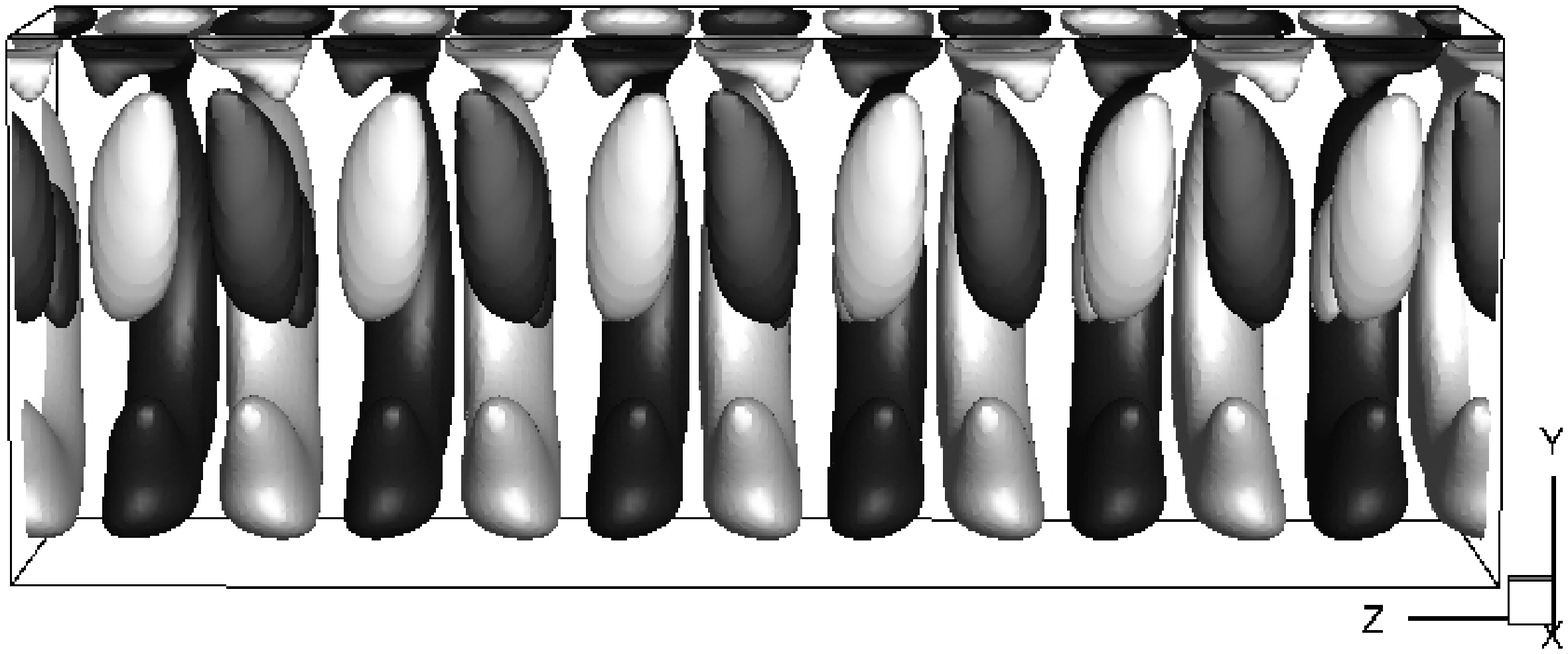}  \\
 \end{tabular}
	\caption{Three views of Taylor-G\"ortler structure, as isosurfaces of dimensionless helicity $hU_q^2/H$ (edge values $\pm 5.8$). Only one half of the spanwise direction is represented for the sake of clarity. Spanwise boundary conditions are periodic. Top pictures are side-views of the structure, slightly seen from above and downstream (flow direction is indicated by the arrow). Bottom picture is a front view from the cavity downstream wall. Black corresponds to negative values of helicity, white to positive values. }
	\label{fig:tgdns}
\end{figure}

\begin{figure}
 \begin{tabular}{c}
	\includegraphics[width=.8\linewidth]{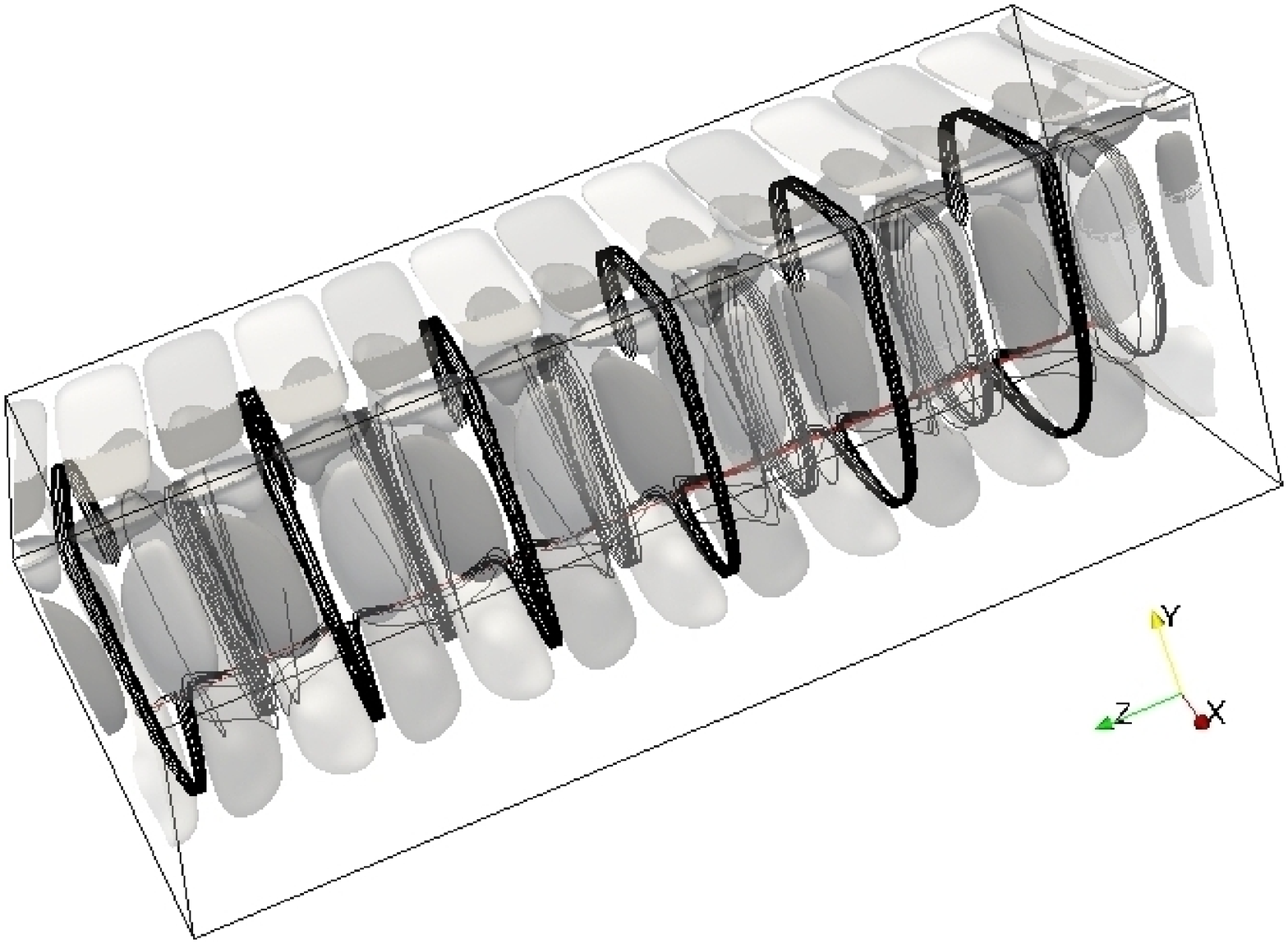} \\ 
	\includegraphics[width=.8\linewidth]{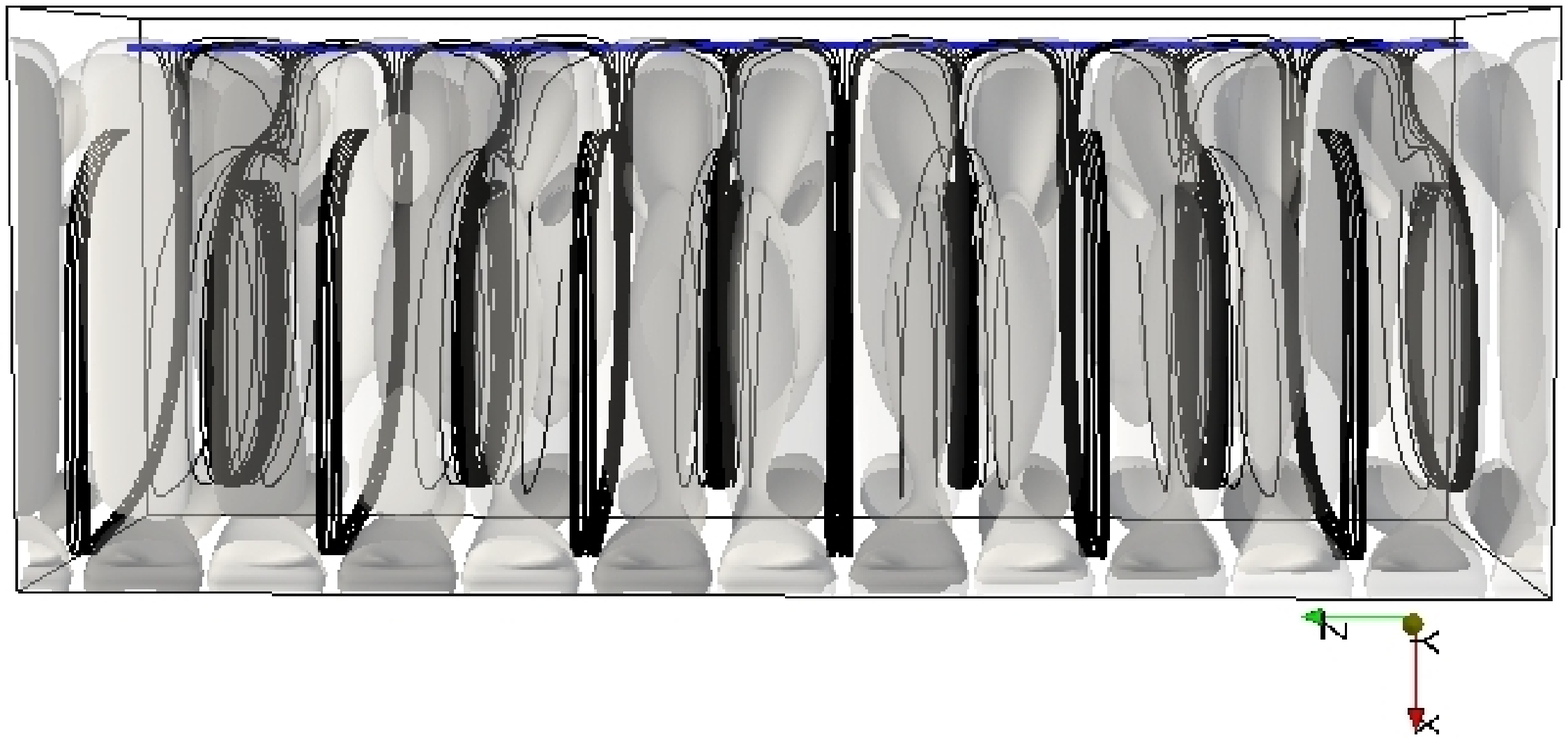} \\ 
 \end{tabular}
	\caption{Two views of streamlines stem from a spanwise line, at the bottom of the cavity, close to the upstream wall (the seeding line is materialised in pictures). Trajectories forward in time form inner loops (gray on top picture), trajectories backward in time form larger loops (black on top picture). Isosurfaces of Figure~\ref{fig:tgdns} are shown in transparency. One hundred particles seed the flow, trajectories are computed by a Runge-Kutta 4 time-integrator.}
	\label{fig:tgtraj}
\end{figure}

Similarly to what is observed in spanwise wall-bounded configurations, raw of Taylor-G\"ortler-like vortical structures take place inside the cavity flow and wind around the main recirculation vortex, as shown in Figure~\ref{fig:tgdns}. However, contrary to the wall-bounded configuration, no structure migration towards the spanwise walls is observed.

Contrary to what could be inferred from helicity representation of Figure~\ref{fig:tgdns}, where tubes of Taylor-G\"ortler vortices appear broken and the sign of helicity changes while moving along the tube, the material tubes are not broken and do form closed tori around the main recirculation (see streamlines of Figure~\ref{fig:tgtraj}). The change of sign in helicity, $h=\vec{\omega }\cdot \vec{v}$, is well understood when noticing that helicity is made of two contributions, $h=\vec{\omega }_\parallel \cdot \vec{v}_\parallel  + \vec{\omega }_\bot \cdot \vec{v}_\bot$, where $\vec{\omega }_\parallel $ is the vorticity component along the tube axis, $\vec{v}_\parallel $ the velocity component parallel to $\vec{\omega }_\parallel $, and $\vec{\omega }_\bot $ and $\vec{v}_\bot$ are the vorticity and velocity components, perpendicular to the tube axis, respectively. While $\vec{\omega }_\parallel $ and $\vec{v}_\parallel$ keep the same orientation all along the Taylor-G\"ortler tube, relatively to each other, $\vec{\omega }_\bot \cdot \vec{v}_\bot$ can change of sign along the tube.

Although the flow is not fully steady, it is insightful to consider streamlines released backward and forward to a seeding line at the upstream bottom quadrant of the cavity (materialized line in Figure~\ref{fig:tgtraj}), and consider streamlines as approximate material trajectories. Trajectories gather into beams and draw funnel-like shapes, materializing the coherent structures highlighted by iso-surfaces of helicity (see Figure~\ref{fig:tgtraj}). In addition, although particles may start aligned in the spanwise direction $z$, their trajectories coil around each other, which reveals an helical flow in Taylor-G\"otler vortices. There is no mass flux between two adjacent counter-rotating vortices. The only spanwise transfer occurs inside both corner vortices, at the bottom of the cavity. 
Particles seeded inside the core of the main recirculation describe concentric circles, characteristic of a solid rotation (not shown in the Figure).

\begin{figure}[t]
	\centering{
	\includegraphics[width=.75\linewidth]{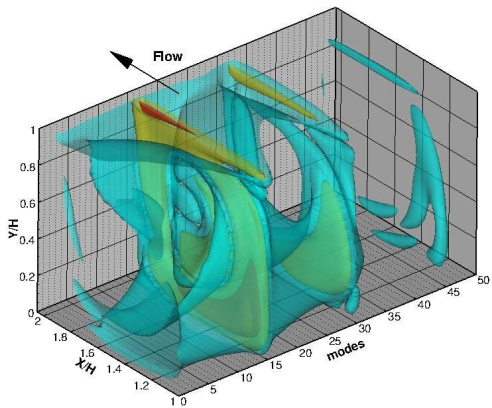} 
	}
	\caption{Each ($x,y$) slice represents the power structure, in the ($x,y$) plane, of spanwise Fourier modes $\mathrm{e}^{ik_mz}$ from a discrete Fourier decomposition of the permanent flow, where $k_m=2\pi m/S$ is the $m^{th}$ mode wave-vector. Slices of fifty modes are shown and connected by isosurfaces with values $1.2u_x/U_q= 10^{-4}$ (blue), $10^{-3}$ (yellow), $10^{-2}$ (red). Most energetic modes are found for $m=15$ and $m=30$ (red colors).}
	\label{fig:tgdnspower}
\end{figure}

\begin{figure}
	\includegraphics[width=.75\linewidth]{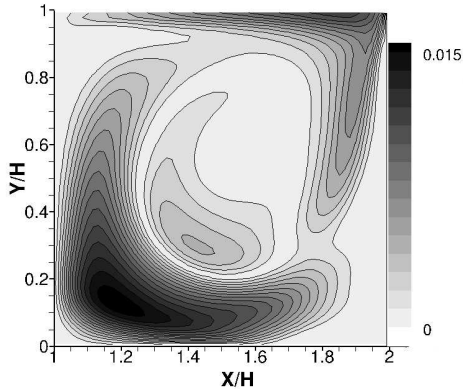} 
 \caption{Streamwise structure of the most energetic spanwise Fourier mode, $m=15$ in Figure~\ref{fig:tgdnspower}. Colorscale encodes power spectral density of $u_x$, in units of $U_q$.}
	\label{fig:tgdnsxy}
\end{figure}

\begin{figure}
	\includegraphics[width=.75\linewidth]{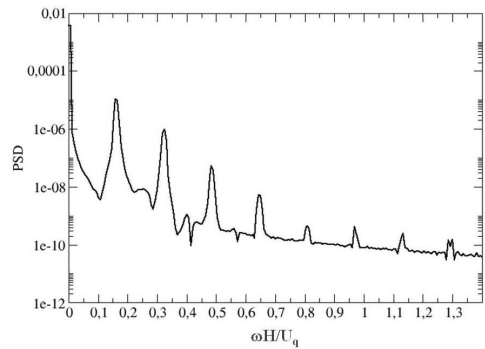} 
	\caption{Power spectral density computed from velocity time-recording at a probe inside the cavity.}
	\label{fig:tgspct}
\end{figure}

A spectral Fourier analysis on velocity components, in the spanwise direction, yields another insight on the spatial arrangement of the flow (see Figure~\ref{fig:tgdnspower}). The most energetic Fourier modes, in space, are found for spanwise wavelengths, $\lambda /H\simeq 0.42$ ($\lambda /S\simeq 0.07$) and $0.84$ ($\lambda /S\simeq 0.14$). Wavelength $\lambda /H=0.42$ is therefore associated with the rake of vortex pairs, while its second harmonic, $\lambda /H=0.84$, is connected to single vortices inside pairs. The streamwise ($x,y$) structure of the most energetic mode is shown in Figure~\ref{fig:tgdnsxy}.

In addition, a spectral Fourier analysis on temporal samplings of velocity, at probes set in the cavity, points out that the flow is not quite stationary and contains a slight oscillating component, as shown in Figure~\ref{fig:tgspct}. This oscillation, found all over the cavity, is characterized by a angular frequency, $\omega H/U_q=0.16$. A flow inspection reveals that this oscillating mode is associated with a swaying motion of Taylor-G\"ortler-rolls, around the main recirculation. In Figure~\ref{fig:tgdns}, the slightly swaying motion is mainly observed on the inner blobs of helicity, close to the upstream cavity wall (better seen on the top right picture).

It is worthwhile noticing that the mode with largest growth-rate found in \cite{BreCol08}, for $\Gamma _L=1$ and Mach numbers $M >0.3$ has a wavenumber $\lambda /H \simeq 0.5$, similar to our $\lambda /H=0.42$. This mode, in \cite{BreCol08}, is expected to be a steady mode ($\omega =0$). In our configuration, the non-linearly saturated regime is found slightly unsteady, with $\omega H/U_q = 0.16$. This discrepancy motivates a three-dimensional linear stability analysis of the flow in the incompressible limit ($M=0$).

\section{Linear stability analysis}

Stability properties of a two-dimensional steady base state, with respect to spanwise perturbations, are now considered. 

\subsection{Principle}

The method is described in \cite{MamTuc95}. It is aimed at characterizing the time-evolution of infinitesimal perturbations, $\boldsymbol{u}'$, with respect to an unstable steady base flow, $\boldsymbol{U}_0$, by means of linearised Navier-Stokes equations. The main instability features are depicted by the leading eigenpairs of the linear evolution operator, namely eigenpairs with the greatest real part. Leading eigenpairs are determined by an Arnoldi method. To shortly remind the procedure, consider the Navier-Stokes equations, linerarized around $\boldsymbol{U}_0$, describing the dynamics of the perturbation:
\begin{equation} 
 \left\{\begin{array}{rcl}
 \displaystyle \frac{\partial{\boldsymbol{u}'}}{\partial t} + \boldsymbol{U}_0\cdot \nabla \boldsymbol{u}' + \boldsymbol{u}'\cdot \nabla \boldsymbol{U}_0 & = & \displaystyle -\nabla p' + \frac{1}{Re}\Delta{\boldsymbol{u}'} \\
 & & \\
 \nabla \cdot\boldsymbol{\boldsymbol{u}'} & = & 0, \\
 \end{array}\right.
 \label{eqn:lin_op}
\end{equation}
In compact form, the system rewrites:
$$
 \frac{\partial \boldsymbol{u}'}{\partial t} = N_U\boldsymbol{u}' + D\boldsymbol{u}' \equiv  A\boldsymbol{u}',
$$
where $N_U$ is the linearised operator of evolution associated with the eulerian part of the Navier-Stokes equations and $D$ is the linear operator associated with viscous terms. Time-evolution of the perturbation is therefore given by:
\begin{equation}
 \boldsymbol{u}_{n+1}'= \exp(A\Delta t)\boldsymbol{u}'_n \equiv B~\boldsymbol{u}'_n,
 \label{eqn:lin_op}
\end{equation}
where $n$ and $n+1$ are the temporal subscripts referring to two consecutive samples with $\Delta t$ apart and $B$ is the discretised linear operator that governs the temporal evolution of $\boldsymbol{u}'$. 
Operators $A$ and $B$ have approximatively the same eigenvectors while their eigenvalues are related by:
\begin{equation} 
 \mu_B =\exp(\mu_A\Delta t).
 \label{eqn:eigval} 
\end{equation}
Therefore, leading eigenvalues $\mu_A$, of $A$, also provide dominant eigenvalues $\mu_B$, of $B$, and reciprocally. Eigenvalues $\mu_B$, and their associated eigenvectors, are determined by considering the Krylov subspace, $U_K$, spanned by $K$ vectors, $\boldsymbol{u}'_k$, the dimension of which is $3N$, $N$ being the number of grid nodes. Dimension $K$ is small and directly related to the number of leading eigenpairs of interest. Vectors $\boldsymbol{u}'_k$ are generated by the same time-stepping procedure as the one used to perform the 3D DNS, with the usual Navier-Stokes equations \eqref{eqn:ns} replaced by their linearised counterpart \eqref{eqn:lin_op}. Time-integration is carried out starting from an initial condition $\boldsymbol{u}'_0$ that will be discussed further. The time stepping procedure runs until the most decaying eigenvectors have vanished. The $K$ vectors $\boldsymbol{u}'_k$ are recorded at a sampling time, $\Delta t$, chosen as a multiple of the numerical time-step, $ \Delta t = \alpha \delta t$. Parameter $\alpha$ is small (usually unity) and adjusted such as to produce results insensitive with respect to the choice of $\Delta t$. Time-integration is not conducted for too long times in order to avoid the solution $\boldsymbol{u}'_k$ to be  completely dominated by the most dominant eigenvector.
The $K$ vectors are ortho-normalised by a Graham-Schmit procedure. The new Krylov subspace is composed of $K$ orthonormalised three components vectors $\boldsymbol{v}_k$, gathered into a $3N \times K$ matrix $V_K$. Moreover, from the $QR$ factorisation of $U_K$, it is possible to determine a relation yielding an approximation of $BV_K$ in the base $V_K$, of the form:
\begin{equation} 
 BV_K = V_KH + r_{K+1},
 \label{eq:arnoldi}
\end{equation}
which is the Arnoldi equation.
Matrix $H$ is a $ K \times K$ Hessenberg matrix, whose elements are the inner products $\left<v_i,Bv_j\right>$. A reference criterion for the selection of dimension $K$ is based on the minimisation of the residual $r_{K+1}$. The diagonalisation of $H$ yields eigenvalues $\nu_k$ and  eigenvectors $\psi_k$. The former approximate the dominant eigenvalues of $B$, and therefore yields leading eigenvalues of $A$ from Eq.~\ref{eqn:eigval}. The later provide the corresponding eigenvectors $\phi_k$ of $B$ and $A$, given by 
$$
 \phi_k=V_K\psi_k.
$$

\subsection{Choice of the base flow and initial perturbation form}

\begin{figure}
	\includegraphics[width=.5\linewidth]{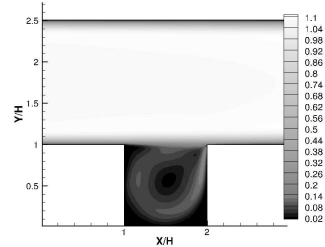}
 \caption{Two-dimensional steady base-flow, solution of equations~\eqref{eqn:ns} for $Re=3\,850$ and $\Gamma _L=1$. }
	\label{fig:baseflow}
\end{figure}

\begin{figure}
 \begin{tabular}{c}
	\includegraphics[width=.75\linewidth]{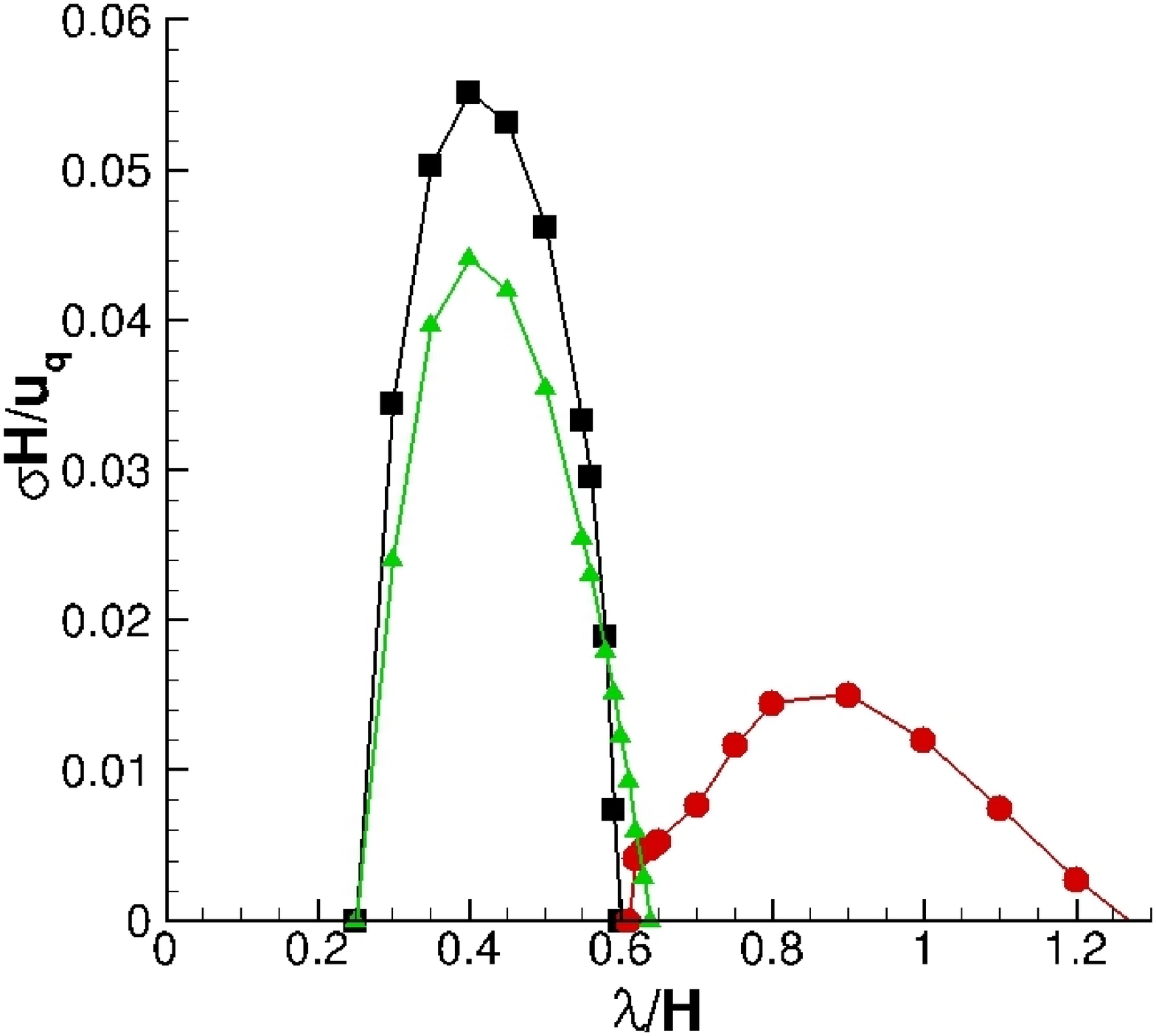} \\
	\includegraphics[width=.75\linewidth]{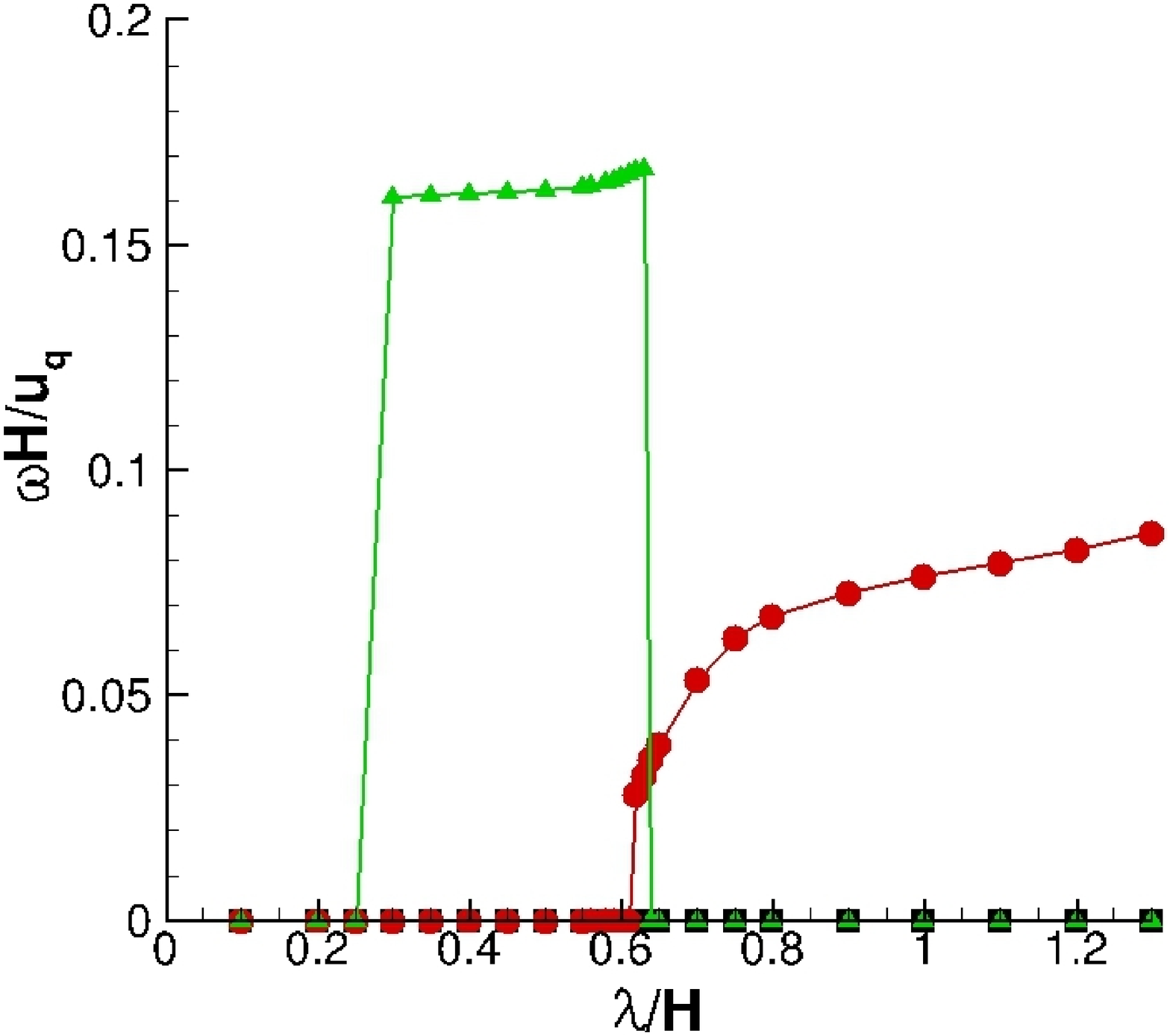} \\
 \end{tabular}
 \caption{Dimensionless growth-rate (top) and angular frequency (bottom) with respect to dimensionless spanwise wavenumber $\lambda /H$. Three families are found: \textit{i)} steady growing modes (black squares), \textit{ii)} oscillatory growing modes, with a constant angular frequency, $\omega H/U_q\simeq 0.16$, and small wavelength, $\lambda /H<0.62$ (green triangles), \textit{iii)} oscillatory growing modes of larger wavelength, with an angular frequency increasing with wavelength (red circles). Mode with the highest growth-rate belongs to family \textit{i)} with a dimensionless wavelength $\lambda /H\simeq 0.4$.}
	\label{fig:tgdnsstab}
\end{figure}

A preliminary two-dimensional numerical simulation, in the streamwise plane ($x,y$), for the same Reynolds number, $Re=3\,850$, exhibits an asymptotically stable steady-state flow, $\boldsymbol{U}_0^{2D}$. As a consequence, in the three-dimensional configuration under study, the instability of the base-flow must occur with respect to spanwise modes, as it is observed in three-dimensional numerical simulations of the Navier-Stokes equations \eqref{eqn:ns}. Therefore, the two-dimensional base-flow, $\boldsymbol{U}_0^{2D}$, is perturbated with initial conditions of the form:

$$ 
 \boldsymbol{u}'(x,y,z)= \boldsymbol{u}_0(x,y) \exp\left(i\frac{2\pi}{\lambda_j}z\right) \exp\left(\mu_jt\right),
$$ 
that is, a mode of wavelength $\lambda_j=L/m$, $m\in \mathbb{N}^\star $, in the spanwise direction, and (complex) growth-rate $\mu_j=\sigma_j +i \omega_j$, with $\sigma _j$ the temporal growth-rate and $\omega _j$ the angular frequency. Only stationary wave-like modes are considered since no spanwise drift is expected in the permanent flow. The governing equations \eqref{eqn:lin_op} rewrite as:
$$
 A(\boldsymbol{u}'_0)=\frac{\partial \boldsymbol{u}'_0}{\partial t} = N_U\boldsymbol{u}'_0 + \left(D_{/(x,y)}-\frac{2\pi}{\lambda_j}\right)\boldsymbol{u}'_0,
$$
where $D_{/(x,y)}$ is the restriction of $D$ to plane ($x,y$).
For each specific wavelength $\lambda_j$, the $K$ leading eigenpairs  ${\mu_{j,k},\phi_{j,k}}$ can be sought out.

\subsection{Linearly growing modes}

\begin{figure}
	\includegraphics[width=.75\linewidth]{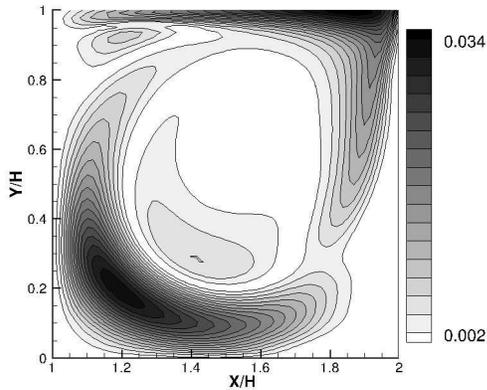} 
 \caption{Streamwise structure (modulus) of the spanwise mode with largest growth rate of the first family ($\omega =0$), referred to as mode\,(i). }
	\label{fig:tgdns_maxgrowthi}
\end{figure}

\begin{figure}
 \begin{tabular}{c}
	\includegraphics[width=.75\linewidth]{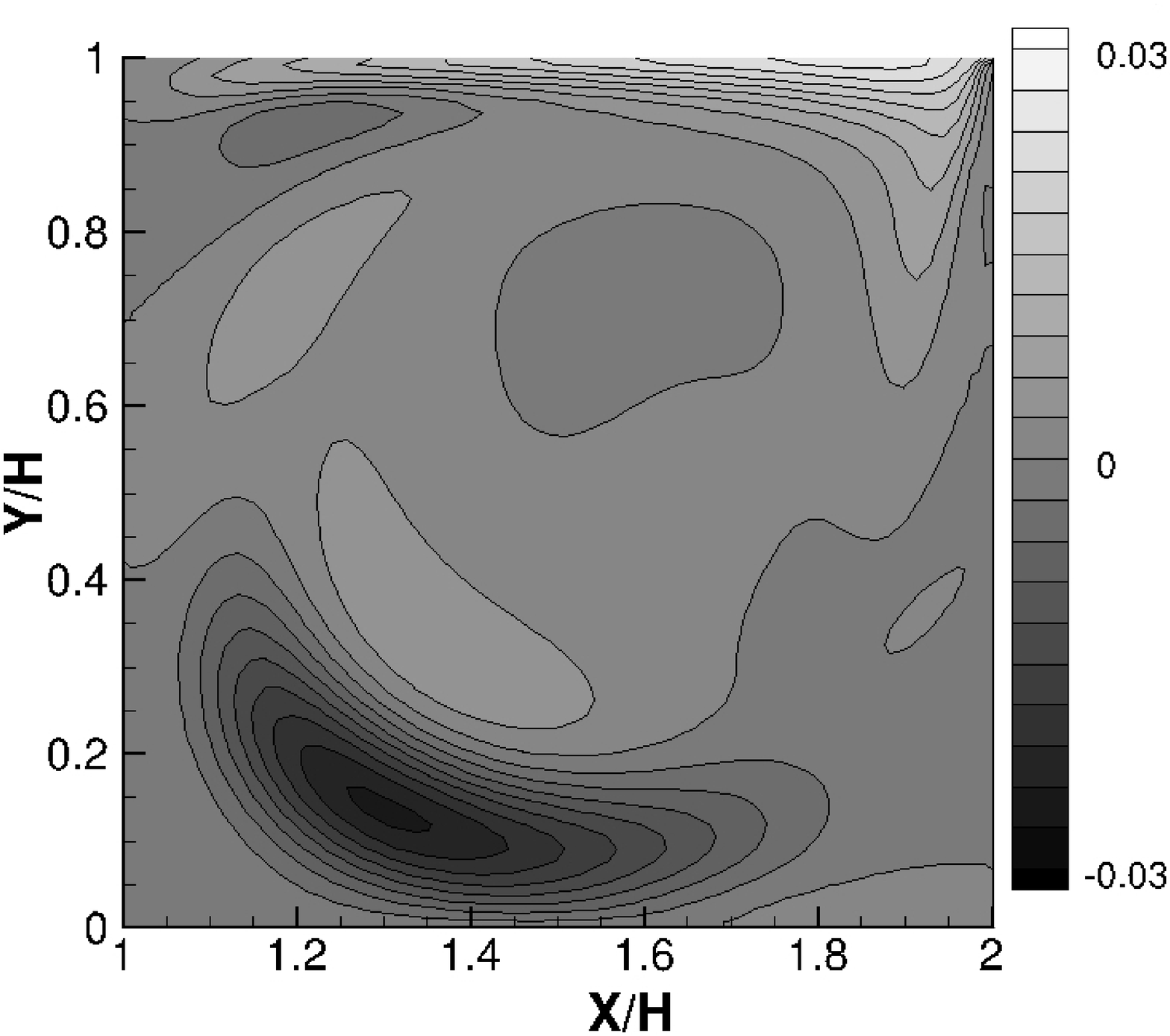} \\
	\includegraphics[width=.75\linewidth]{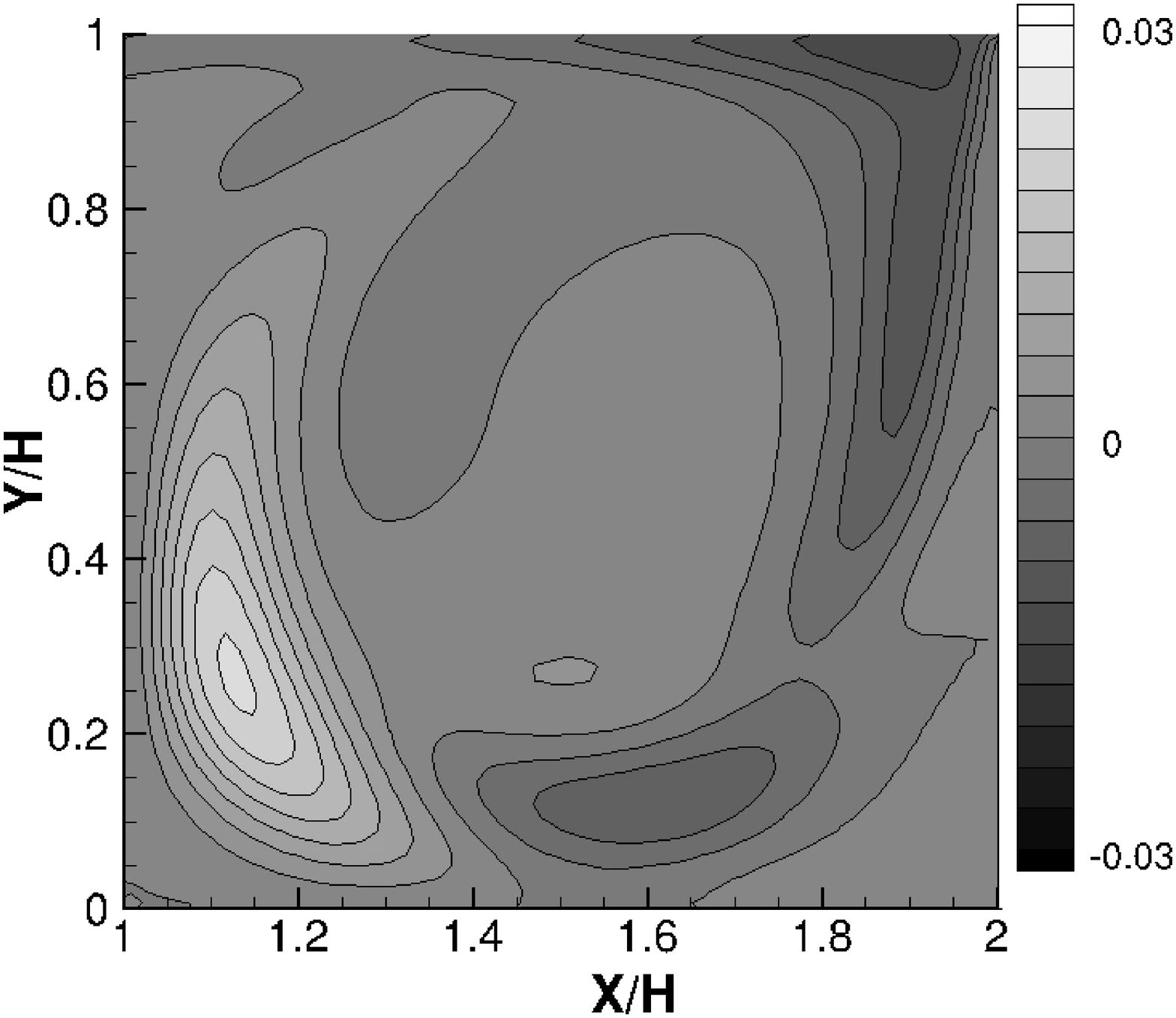} \\
 \end{tabular}
 \caption{Streamwise structure of the spanwise eigen-function with largest growth-rate of the second family ($\omega H/U_q=0.16$), referred to as mode\,(ii): real part on the top, imaginary part on the bottom. The colorscale encodes $u_x$, in units of $U_q$.}
	\label{fig:tgdns_maxgrowthii}
\end{figure}

Linear stability analysis is carried out for a set of initial perturbations $\lambda /H$ in the range 0 to about $3/2$, corresponding to $S/4$. As a first result, it is found that spanwise modes with vanishing wavenumbers have negative growth-rate. Consequently, the two-dimensional steady base-flow is unstable with respect to spanwise, rather than purely streamwise, instabilities. 
Three branches of instability are found, whose temporal growth-rate $\sigma $ and angular frequency $\omega$, as functions of $\lambda$, are illustrated in Figure~\ref{fig:tgdnsstab}. One branch (black squares) is associated with a stationary bifurcation ($\omega =0$), on the wavelength range $\lambda /H\in [0.24, 0.618]$. The largest growth-rate on this branch is found at $\lambda /H=0.402$, close to the wavelength found for the raw of Taylor-G\"ortler vortices ($\lambda _{TG}/H\simeq 0.42$). We will refer to this mode as mode\,(i). A spatial representation of mode\,(i) is shown in Figure~\ref{fig:tgdns_maxgrowthi}.

In Figure~\ref{fig:tgdnsstab} is also found, over the range of wavelength $\lambda /H\in [0.24,0.642]$, another branch of solutions (green triangles) with positive growth-rates, and angular frequencies roughly constant over the range, at $\omega _{ii} H/U_q=0.16$, very close to the one of Taylor-G\"ortler vortices. For this branch, the highest growth-rate is found, again, for $\lambda /H\simeq 0.4$, and we will refer to this mode as mode\,(ii).  A comparison of both real and imaginary parts of mode\,(ii), in Figure~\ref{fig:tgdns_maxgrowthii}, reveals a swaying motion of the structure around the main recirculation.

When the growth-rate of the stationary branch crosses zero, at $\lambda /H=0.618$, a new branch rises, on the range $\lambda/H\in [0.618,1.32]$, characterized by a non-zero angular frequency (red circles). Following this oscillatory branch, the angular frequency increases from $0$ to about $\omega H/U_q=0.092$, with a linear variation beyond $\lambda /H=0.78$. In this family, the mode with the largest growth-rate, referred to as mode\,(iii), is found for $\lambda _{iii}/H\simeq 0.82$ with $\omega _{iii} H/U_q\simeq 0.075$.

Finally, no spanwise mode is found with positive growth-rate when $\lambda /H>1.32$.

\subsection{From instability to permanent flow}

The inner flow most energetic structure, in the permanent flow, has both a spanwise wavelength, $\lambda _{TG}/H$, and a streamwise structure, in close accordance with those of mode\,(i), the most linearly unstable mode in our linear stability analysis (compare Figures~\ref{fig:tgdnsxy} and \ref{fig:tgdns_maxgrowthi}). However, mode\,(i) is steady, whereas the permanent regime exhibits a slightly unsteady swaying motion. 
Note that transient dynamics, although not fully excluded, are very unlikely, since simulations in the permanent flow cover a time-range of $1\,440$ time-units $H/U_q$. 
In fact, when considering the power spectral density of Figure~\ref{fig:tgspct}, in the permanent regime, a main peak is found  at $\omega _{TG}H/U_q\simeq 0.16$, which precisely is the angular frequency of modes of the second family. Moreover, the streamwise structure of mode\,(ii), in that family, exhibits a swaying-like motion, as shown in Figure~\ref{fig:tgdns_maxgrowthii}, when comparing both real and imaginary parts. Such features suggest that this mode, as mode\,(i), is also selected by the flow.  
Note that linear modes are recovered in the permanent regime without significant distortion, despite of non-linear effects. 

It is worth to note that the angular frequency of the third family mode, with the highest growth-rate, is $\omega _{iii}H/U_q\simeq 0.07$. Would such a mode be present in the flow, it could couple to mode\,(ii) and generate components at $(\omega _{ii}\pm \omega _{iii})H/U_q$, i.e. 0.09 and 0.23. In the power spectral density of Figure~\ref{fig:tgspct}, a lobe is actually found around $\omega H/U_q\simeq 0.25$, suggesting a non-linear (quadratic) coupling of mode\,(iii) with mode\,(ii). 
In addition, the spanwise Fourier mode associated with $m=30$, in Figure~\ref{fig:tgdnspower} for the permanent regime, is not found in the linear stability analysis, indictaing that this mode is rather generated by quadratic non-linear self-interaction of mode $m=15$.

\section{Conclusion}

Three dimensional direct numerical simulations of an open square cavity flow are performed in the incompressible limit and at Reynolds number $Re=3\,850$, below the threshold of shear layer instability. They reproduce most of the inner flow characteristic features observed in experiments. As a result of centrifugal instabilities, vortical structures develop and form spanwise alley of pairs of counter-rotating vortices \cite{Faure2007,Faure2009}. 
When spanwise boundary conditions are periodic, the alley of vortices is fixed. 
However, in the non-linearly saturated regime, a closer inspection of the flow reveals a slightly swaying motion of the structure of vortices around the main recirculation flow.

The fastest (linearly) growing spanwise mode, in a linear stability analysis of the steady base state, has a wavelength $\lambda /H=0.4$, fairly close to the wavelength associated with the spanwise alley of vortices. However, from stability analysis, this mode is expected stationary. Yet, the swaying motion is caught by a mode of the second family, whose growth-rate is maximal for wavelengths of the order of $\lambda _{TG}$. Surprisingly, streamwise structures, wavelength and angular frequencies of linear modes are not much distorted in the non-linear regime. Non-linear effects are revealed by coupling modes of two families, namely the modes with largest growth-rates from families (ii) and (iii). 

Note that beyond the drift motion initiated by the pumping effect at the spanwise walls, a secondary bifurcation towards an unsteady oscillating phenomenon of the alley of vortices has been reported in experiments, see for instance \cite{Faure2009}. Although the streakline features of this secondary instability could possibly be understood as the result of a swaying motion, it is not possible, yet, to ascertain whether both phenomenons, in simulations and experiments, be related or not.

\section*{Aknowledgments}

This work has been supported by DIGITEO project FLUCTUS.
J.B. gratefully acknowledges fruitful discussions with J. de Vicente Buendia.
The authors wish to thank Th.M.~Faure for his most helpful advice.

\end{document}